\documentclass[notitlepage,12pt]{revtex4-1}

\usepackage{color}

\usepackage[utf8]{inputenc}
\usepackage{amsmath,amssymb,amstext}
\usepackage{amsthm}
\usepackage{dsfont}
\usepackage{graphicx}

\def\sout{\bgroup\markoverwith
{\textcolor{red}{\rule[0.5ex]{2pt}{0.5pt}}}\ULon}
\def\be{\begin{equation}}
\def\ee{\end{equation}}
\def\bes{\begin{equation*}}
\def\ees{\end{equation*}}
\def\bea{\begin{eqnarray}}
\def\eea{\end{eqnarray}}
\def\beas{\begin{eqnarray*}}
\def\eeas{\end{eqnarray*}}
\def\bal#1\eal{\begin{align}#1\end{align}}
\def\bals#1\eals{\begin{align*}#1\end{align*}}
\newcommand{\bra}[1]{\langle #1|}
\newcommand{\ket}[1]{|#1\rangle}
\newcommand{\braket}[2]{\langle #1|#2\rangle}

\renewcommand{\vec}{\vectorsym}

\renewcommand*{\vec}[1]{\boldsymbol{#1}}

\usepackage[normalem]{ulem}

\usepackage{braket}

\graphicspath{{figures/}}

\makeatletter
\AtBeginDocument{\let\LS@rot\@undefined}
\makeatother

\begin{document}

\title{Hamiltonian for the Hilbert-Pólya Conjecture}

\author{Enderalp Yakaboylu}
\email{yakaboylu@proton.me}

%\email{enderalp.yakaboylu@uni.lu}
%\affiliation{Max Planck Institute of Quantum Optics, 85748 Garching, Germany}
%
%\affiliation{Department of Physics and Materials Science, University of Luxembourg, L-1511 Luxembourg}

\date{\today}

\begin{abstract}

We introduce a Hamiltonian to address the Hilbert-Pólya conjecture. The eigenfunctions of the introduced Hamiltonian, subject to the Dirichlet boundary conditions on the positive half-line, vanish at the origin by the nontrivial zeros of the Riemann zeta function. Consequently, the eigenvalues are determined by these nontrivial Riemann zeros. If the Riemann hypothesis (RH) is true, the eigenvalues become real and represent the imaginary parts of the nontrivial zeros. Conversely, if the Hamiltonian is self-adjoint, or more generally, admits only real eigenvalues, then the RH follows. In our attempt to demonstrate the latter, we establish the existence of a similarity transformation of the introduced Hamiltonian that is self-adjoint on the domain specified by an appropriate boundary condition, which is satisfied by the eigenfunctions through the vanishing of the Riemann zeta function. Our result can be extended to a broader class of functions whose zeros lie on the critical line.

\end{abstract}

\maketitle

\section*{Introduction}

The Riemann hypothesis (RH) is considered to be one of the most important unsolved problems in mathematics, if not the most. The hypothesis states that every nontrivial zero of the Riemann zeta function, denoted by $\rho$, lies on the critical line $\text{Re}(\rho) = 1/2$. The Riemann zeta function, which is defined as follows
\bes
\zeta(s) = \sum_{m=0}^\infty \frac{1}{(m+1)^{s}} 
\ees
for $\text{Re}(s) > 1$, can be extended for other complex values of $s$ via analytic continuation~\cite{titchmarsh1986theory}. Particularly, for $\text{Re}(s) > 0$ except for the points where $(1 - 2^{1-s})$ is zero (which we consider hereafter), it can be expressed as $\zeta(s) =  (1 - 2^{1-s})^{-1} \eta(s)$ via the Dirichlet eta function:
\bes
 \eta(s) =  \sum_{m=0}^\infty \frac{(-1)^m}{(m+1)^s} = \frac{1}{\Gamma(s)} \int_0^\infty d u \, \frac{u^{s-1}}{1+ e^{u}} \, .
\ees

Despite strong numerical evidence of its validity - it is known to be true for the first $10^{13}$ zeros~\cite{gourdon20041013}, and various analytical attempts, the hypothesis has not yet been proven. Among several pure mathematical approaches to establishing the RH, there is one particular approach that has also attracted physicists; the Hilbert-Pólya conjecture. It asserts that the imaginary parts of the nontrivial Riemann zeros correspond to eigenvalues of a self-adjoint Hamiltonian. More precisely, there exists a self-adjoint Hamiltonian, $\hat{H}$, such that the eigenvalues of the operator $(1/2 + i \hat{H})$ coincide with all the nontrivial zeros of the Riemann zeta function. The self-adjoint property of the Hamiltonian would imply that its eigenvalues are real and hence the RH follows. Consequently, the mere existence of such a Hamiltonian, whose eigenvalues are all real, would prove the RH. Although attempts to construct such a Hamiltonian have been unsuccessful to date, the Hilbert-Pólya conjecture has gained firmer grounding through the contributions of various seminal works~\cite{selberg1956harmonic,montgomery1973pair,
odlyzko1987distribution,okubo1998lorentz,connes1999trace,berry1999h,berry1999riemann,
sierra2007h,sierra2019riemann,bender2017hamiltonian,
bellissard2017comment,das2019supersymmetry}.  One of the most significant works among them is the Berry-Keating program~\cite{berry1999h,berry1999riemann}.

Berry and Keating conjectured that the classical limit of such a Hamiltonian would take the form $x p$, which was also considered by Connes~\cite{connes1999trace}. This simple suggestion was based on a rather heuristic and semiclassical analysis. As a quantum counterpart, they considered the simplest  formally self-adjoint Hamiltonian,
\bes
\hat{H}_\text{BK} = \frac{1}{2}\left( \hat{x} \, \hat{p}  + \hat{p} \,  \hat{x} \right) \, .
\ees
Although the Berry-Keating Hamiltonian corresponds to the correct structure of the eigenvalues, which was previously obtained by Okubo~\cite{okubo1998lorentz} in a different Hamiltonian, and provides the average number of the nontrivial zeros up to a given height, this Hamiltonian is still far from being concrete. Especially, it is unclear how to reveal the Riemann zeta function in the corresponding eigenvalue equation, or on which domain this Hamiltonian should act to be self-adjoint, given the imposed boundary condition for the vanishing of the Riemann zeta function. 

Essentially, the Hilbert-Pólya conjecture involves two stages: (I) finding an operator whose eigenvalues correspond to the imaginary parts of the nontrivial Riemann zeros, assuming the RH is correct; and (II) proving that this operator is self-adjoint or, more generally, that it admits only real eigenvalues. The latter is the most challenging stage of the Hilbert-Pólya conjecture, as it requires imposing appropriate boundary conditions such that the operator is self-adjoint on the corresponding domain. Indeed, while some operators proposed in the past satisfy the first stage (see for instance Ref.~\cite{bender2017hamiltonian}), the second stage has remained elusive up to the present, which we now endeavor to address.

Within this manuscript, we introduce a Hamiltonian that could potentially satisfy both of these stages. Particularly, in Sec.~\ref{sec_ham}, corresponding to stage (I), we show that the eigenvalues of the introduced Hamiltonian are given in terms of the nontrivial Riemann zeros. Furthermore, if the RH holds true, then the eigenvalues are real and correspond to the imaginary parts of the nontrivial zeros. Subsequently, in Sec.~\ref{sec_rh}, we investigate whether the introduced Hamiltonian has only real eigenvalues, thereby establishing its viability as an approach to the RH within the Hilbert-Pólya conjecture. Accordingly, we attempt to demonstrate the existence of a similarity transformation that maps the introduced Hamiltonian to a self-adjoint Hamiltonian. The corresponding domain is identified by an appropriate boundary condition, which the eigenfunctions satisfy through the vanishing of the Riemann zeta function, thereby addressing stage (II). We conclude the paper with a discussion of extending our findings to a wider class of functions whose zeros are only on the critical line. Additional technical details can be found in Appendix.

\section{Hamiltonian} \label{sec_ham}

Building upon the earlier works we referenced, we introduce the following Hamiltonian for the Hilbert-Pólya conjecture
\be
\label{initial_ham}
\hat{H} = \hat{S} \, \hat{H}_\text{BK}  \, \hat{S}^{-1}  \quad \text{with}\quad \hat{S} = t^{\hat{N}} \frac{e^{\alpha \hat{x}}}{1 + e^{\hat{x}}} \, ,
\ee
acting on the Hilbert space $L^2[0,\infty)$. The Hamiltonian is basically a similarity transformation of the Berry-Keating Hamiltonian, involving the position operator and the number operator defined on the positive half-line:
\bes
\hat{N} = \frac{1}{2}\left(\hat{x} \, \hat{p}^2 + \hat{p}^2 \, \hat{x} + \frac{\hat{x}}{2} \right) \, .
\ees
Moreover, we define $\alpha = (1+t)/(2-2t)$ with $\text{Re}(t)<1$. 

In the proposed Hamiltonian~\eqref{initial_ham}, the Berry-Keating Hamiltonian is well-defined and self-adjoint on the domain $\{\phi(x) \in L^2[0,\infty); \,  \phi(x) \ \text{abs. cont. on} \ [0,\infty) \}$, which can be shown by comparing the dimensions of the called ``deficiency subspaces'' (both the deficiency indices are equal to zero)~\cite{bonneau2001self,faris2006self,twamley2006quantum}. Furthermore, on the same domain it is the generator of the dilation operator, $ e^{ i \lambda \hat{D}} $ (henceforth we refer to the Berry-Keating Hamiltonian as $\hat{D} = \hat{H}_\text{BK} $). This means that $\hat{D}$ is unitarily equivalent to a multiplication operator, which is a formulation of the spectral theorem for self-adjoint operators~\cite{reed1981functional}. The dilation operator appears in conformal field theories as a conformal transformation and in quantum optics as the squeeze operator, where the parameter $\lambda$ labels the amount of squeezing. The operator possesses the following properties;
\bes
e^{i \lambda \hat{D}} \, \hat{p} \, e^{-i \lambda \hat{D}}  = e^{-\lambda} \, \hat{p}\, ,\quad e^{ i \lambda \hat{D}} \, \hat{x} \, e^{-i \lambda \hat{D}} = e^{\lambda} \, \hat{x} \, , \ \text{and} \  \bra{x} e^{i \lambda \hat{D}} \ket{\psi}  = e^{\lambda/2} \psi (e^{\lambda} x) \, .
\ees

The eigenvalue differential equation for the Berry-Keating Hamiltonian is expressed as, 
\bes
\bra{x} \hat{D} \ket{\phi_s} = \left( - i x \frac{d}{d x} - \frac{i}{2} \right) \, \phi_s (x) = \mathcal{E}_s  \, \phi_s (x) \, ,
\ees
which can be solved by employing $ \phi_s (x) \equiv \braket{x| \phi_s}  = x^{s-1}/\sqrt{2 \pi} $, leading to the eigenvalues $\mathcal{E}_s  = i (1/2 - s)$. Given that the eigenvalues of a self-adjoint Hamiltonian must be real, the parameter $s$ is necessarily a complex number with $\text{Re}(s) = 1/2$. This requirement implies that $  \phi_s (x) $ act as generalized eigenfunctions for the cases where $\text{Re}(s) = 1/2$, and the imaginary part, $\text{Im}(s)$, serves as a quantum number that identifies the associated eigenvalues $\mathcal{E}_s $. While the form of the eigenvalues provides a clue to the validity of the Hilbert-Pólya conjecture, the challenging part of the problem lies in identifying these eigenvalues in terms of the nontrivial Riemann zeros.

The eigenvalue equation for the number operator, on the other hand,
\bes
\bra{x} \hat{N} \ket{n} = \left(- x \frac{d^2}{d x^2} - \frac{d}{d x} + \frac{x}{4} \right) \,  \chi_n (x) = \varepsilon_n \, \chi_n (x) \, ,
\ees
can be recognized as the well-established Laguerre ODE with the eigenfunctions $  \chi_n (x) \equiv \braket{x|n}= e^{-x/2} L_n (x)$ and the eigenvalues $\varepsilon_n = (n + 1/2)$, where $n = 0 , 1, \cdots \infty$~\cite{arfken2011mathematical}. To be more precise, the operator $\hat{N}$ is the shifted number operator. The self-adjointness of the number operator on the Hilbert space $L^2[0,\infty)$ can be shown within the framework of Sturm-Liouville theory~\cite{teschl2014mathematical}.

As the Hamiltonian~\eqref{initial_ham} is a similarity transformation of the Berry-Keating Hamiltonian, the eigenvalues of the introduced Hamiltonian can be represented as $\mathcal{E}_s$, while the associated eigenstates can be characterized by
\bes
\ket{\Psi_s} = \hat{S} \ket{\phi_s}  \, .
\ees

The corresponding eigenfunctions in the position space, denoted as $ \braket{x|\Psi_s} \equiv \Psi_s (x) $, are defined as the solutions to the eigenvalue differential equation for the Hamiltonian~\eqref{initial_ham}, with specific boundary conditions imposed. These eigenfunctions can be expressed as
\bal
\label{eigen_func}
\nonumber \Psi_s (x)  & = \int_0^\infty dz \, \bra{x}\hat{S} \ket{z} \braket{z|\phi_s} = \int_0^\infty \frac{dz}{\sqrt{2\pi}}\, \frac{z^{s-1} \, e^{\alpha z}}{1 + e^{z}}  \bra{x} t^{\hat{N}} \ket{z} \\
& = \frac{\sqrt{t}}{\sqrt{2\pi}}\int_0^\infty dz\, \frac{z^{s-1} \, e^{\alpha z}}{1 + e^{z}}  \sum_{n=0}^\infty t^n \chi_n (x) \chi_n (z) \, ,
\eal
which vanish in the limit of $x\to \infty$. Following that, we impose the Dirichlet boundary condition at the origin, $\Psi_s (0) = 0$, to specify the possible values of the eigenvalues, $\mathcal{E}_s$, associated with the eigenfunctions, $\Psi_s (x)$. The motivation of such a boundary condition simply follows from the Hilbert space $L^2[0,\infty)$. The eigenfunctions at the boundary are, then, given by
\bes
\Psi_s (0) = \frac{\sqrt{t}}{\sqrt{2\pi}}\int_0^\infty dz\, \frac{z^{s-1} \, e^{\alpha z}}{1 + e^{z}}  \sum_{n=0}^\infty t^n \chi_n (z) \, ,
\ees
where we employed $\chi_n(0) = 1$. The above series in the integral is recognized as the generating function of Laguerre polynomials,
\bes
\label{genereting_func}
e^{-z/2}\sum_{n=0}^\infty t^n L_n (z)  =  \frac{e^{-z/2} \, e^{- t z/ (1-t)} }{1-t} =  \frac{e^{- \alpha z} }{1-t} \, .
\ees
We note that the series converges to the right hand side of the equal sign for $|t| < 1$. Nevertheless, it can be analytically continued to all $ \text{Re}(t) < 1$ via the Borel sum, see Appendix~A. By further using the expression of the Riemann zeta function via the Dirichlet eta function, we identify the eigenfunctions at the boundary as
\be
\label{bc_zeta} \Psi_s (0) = \frac{1}{\sqrt{2\pi}} \frac{\sqrt{t}}{1-t} (1-2^{1-s}) \Gamma(s) \zeta(s) \, .
\ee
The Dirichlet boundary condition can only be satisfied by the vanishing of the Riemann zeta function, $\zeta(s)  = 0$. Therefore, the eigenvalues of the Hamiltonian~\eqref{initial_ham} are expressed in terms of the nontrivial Riemann zeros, $\rho$, as $\mathcal{E}_s = \left. i (1/2 -s) \right|_{s=\rho}$.

\section{Self-adjointness} \label{sec_rh}

Up to this point, we have shown that the eigenvalues of the Hamiltonian~\eqref{initial_ham} are identified in terms of the nontrivial zeros of the Riemann zeta function. If the RH holds true, then the eigenvalues become real and given by the imaginary parts of the nontrivial Riemann zeros, $\mathcal{E}_s = \text{Im}(\rho)$. This raises the question of whether the introduced Hamiltonian can be utilized to establish the RH within the scope of the Hilbert-Pólya conjecture. This would require proving that the Hamiltonian~\eqref{initial_ham} admits only real eigenvalues.

Before we attempt to demonstrate the reality of all the eigenvalues, we would like to highlight the following point. If we chose $t = e^{i \theta}$ with $\theta \in \mathbb{R}$ (excluding $t = 1$), the transformation $\hat{U} = t^{\hat{N}} e^{\alpha \hat{x}}$ would become unitary. Consequently, the Hamiltonian would be expressed as a unitary transformation of the operator: $(1 + e^{\hat{x}})^{-1} \hat{H}_\text{BK} (1 + e^{\hat{x}})$. Furthermore, as the latter is self-adjoint on the Hilbert space of all square-integrable functions with respect to the weight function $(1 + e^x)^2$ on the interval $[0, \infty)$, the introduced Hamiltonian would also be self-adjoint on the same domain.
However, it would be uncertain if the Hamiltonian remains self-adjoint once the Dirichlet boundary condition is imposed. This is likely the most challenging aspect of the Hilbert-Pólya conjecture, as boundary conditions might restrict the domain in a way that the given operator may no longer be self-adjoint. Therefore, for the rest of the manuscript, we focus on the domain determined by the imposed boundary condition, $\Psi_s (0) = 0$, while continuing to consider $\text{Re}(t) < 1$ in general. 

First of all, instead of the Hamiltonian~\eqref{initial_ham}, we consider its unitarily equivalent form given by
\be
\label{initial_ham_trans}
\hat{\widetilde{H}} = e^{i\lambda \hat{D}} \hat{H} e^{-i \lambda \hat{D}} \, , 
\ee
where $\lambda$ is a finite real parameter, and we denote the corresponding eigenstates as $\ket{\widetilde{\Psi}_s} = e^{i\lambda \hat{D}} \ket{\Psi_s}$. We note that under such a unitary transformation, the Dirichlet boundary condition remains the same, i.e.,  $\widetilde{\Psi}_s (0) = \Psi_s(0) =  0$, which follows from $\widetilde{\Psi}_s(x) = \bra{x} e^{i\lambda \hat{D}} \ket{\Psi_s} = e^{\lambda/2} \Psi_s(e^\lambda x)$.

The Hamiltonian~\eqref{initial_ham_trans} can be expressed as
\be
\label{half_trans_ham}
\hat{\widetilde{H}} =e^{i \lambda \hat{D}} t^{\hat{N}} \left( \hat{D} + i \alpha \hat{x} - i \sum_{m=0}^\infty c_m \hat{x}^m \right)  t^{-\hat{N} } e^{-i \lambda \hat{D}} \, .
\ee
Here we used the series expansion of $x (1+e^{-x})^{-1} = \sum_{m=0}^\infty c_m x^m $, where $c_m = B_m(2^m-1)/m! $ and $B_m$ are the Bernoulli numbers with the convention $B_1 = +1/2$. We note that the sum converges only for $ 0< |x| < \pi $. Nevertheless, the series is Borel summable for all $\text{Re}(x)>0$ and $\text{Im}(x)< \pi $, see Appendix~A.

As a next step, we introduce the following formally self-adjoint operator
\bes
\hat{K} = \hat{N} - \frac{\hat{x}}{2} =  \frac{1}{2}\left(\hat{x} \, \hat{p}^2 + \hat{p}^2 \, \hat{x} - \frac{\hat{x}}{2} \right) \, .
\ees
It is straightforward to show that the three operators, $\hat{D}$, $\hat{N}$, and $\hat{K}$, fulfill the $su(1,1)$ algebra:
\be
\label{su11}
[\hat{D},\hat{N}] = i \hat{K} \, ,\quad  [\hat{N}, \hat{K}] = i \hat{D} \, , \quad [\hat{K}, \hat{D}] = -i \hat{N} \, .
\ee
The Casimir operator for this algebra is given by $ \hat{C} = \hat{D}^2 + \hat{K}^2 - \hat{N}^2 = \hat{I}/4 $, with $\hat{I}$ being the identity operator, see Refs.~\cite{perelomov1977generalized,barut1971new}. We further define the ladder operators $\hat{N}_\pm = \hat{K} \pm i \hat{D}$ such that $ [\hat{N}, \hat{N}_\pm] = \pm \hat{N}_\pm $ and $ [\hat{N}_+ , \hat{N}_-] = -2 \hat{N} $. These commutation relations together with the Casimir operator allow us to identify the action of the ladder operators on the eigenstates of the number operator; $ \hat{N}_+ \ket{m} = (m+1) \ket{m+1} $ and $ \hat{N}_- \ket{m} = m \ket{m-1} $.

Then, by using $\hat{x} = 2(\hat{N} - \hat{K})$ and the $su(1,1)$ algebra~\eqref{su11}, the Hamiltonian~\eqref{half_trans_ham} can be written as
\bal
\label{ham_num0}
\nonumber \hat{\widetilde{H}} & = \hat{D}  \frac{t^2+1}{2 t}- i \left( \hat{K} \text{ch} \lambda - \hat{N} \text{sh} \lambda  \right) \frac{t^2-1}{2 t} \\
& + 2 i \alpha \left(  \hat{N} \left( \text{ch} \lambda +   \frac{t^2+1}{2 t} \text{sh} \lambda  \right)  - \hat{K} \left(\text{sh} \lambda  +  \frac{t^2+1}{2 t} \text{ch} \lambda \right)  - i \hat{D}  \frac{t^2-1}{2 t} \right) \\
\nonumber  &  - i \sum_{m=0}^\infty c_m 2^m \left(  \hat{N} \left( \text{ch} \lambda +   \frac{t^2+1}{2 t} \text{sh} \lambda  \right)  - \hat{K} \left( \text{sh} \lambda  +  \frac{t^2+1}{2 t} \text{ch} \lambda \right)  - i \hat{D}  \frac{t^2-1}{2 t} \right)^m \, .
\eal
If we specify $\text{sh} \lambda = 2t/ (t^2 -1)$ with $t \in  \mathbb{R}$ (excluding $t^2=1$), the above Hamiltonian simplifies to 
\be
\label{ham_num}
\hat{\widetilde{H}} = i \hat{N} -i \hat{N}_- - i \sum_{m=0}^\infty c_m  \left(\beta \hat{N}_- \right)^m \, ,
\ee
where $\beta = (t^2-1)/t$ and the corresponding series is Borel summable under the condition that $\text{Re}(\beta)  > 0$. Accordingly, without loss of generality, we opt to set $t$ such that $\beta = 1$.

The eigenvalue equation associated with the Hamiltonian~\eqref{ham_num} can be solved by first expanding the eigenstates of the Hamiltonian in the basis of the eigenstates of the number operator: $\ket{\widetilde{\Psi}_s} = \sum_{n=0}^\infty f_n \ket{n}$, and then projecting onto a particular eigenstate $\ket{k}$. Accordingly, the eigenvalue equation for the Hamiltonian~\eqref{ham_num} is expressed as
\be
\label{eigenvalue_eq}
 i f_k \left(k+\frac{1}{2} \right) -i f_{k+1}(k+1) - i \sum_{m=0}^\infty c_m f_{k+m} \frac{(k+m)!}{k!} = \mathcal{E}_s \, f_k \, .
\ee
The form of the eigenvalue equation~\eqref{eigenvalue_eq} suggests that $f_k = g_k / k!$. Furthermore, if we define $g_k$ in terms of the Mellin transform as 
\bes
g_k = \{\mathcal{M} g_s \}(k) = \int_0^\infty dz \, z^{k-1} g_s (z) \, ,
\ees 
then the eigenvalue equation~\eqref{eigenvalue_eq} can be formulated as a differential equation for $g_s(z)$  in the called Mellin space
\bes
\label{diff_eq}
\left( - i z \frac{d}{d z} + \frac{i}{2} -i z - \frac{i z}{1+e^{-z}} \right) g_s(z) = \mathcal{E}_s \,  g_s (z) \, .
\ees
The solutions are given by 
\be
\label{sol_in_mellin}
g_s(z)=g_0 (s) \, z^s \frac{e^{-z}}{1+e^{z}} \, ,
\ee
with the normalization coefficient $g_0 (s)$, which correspond to the eigenfunctions~\eqref{eigen_func} via
\bes
\widetilde{\Psi}_s (x) = \sum_{n=0}^\infty \frac{\chi_n (x)}{n!} \int_0^\infty dz\, z^{n-1} g_s (z) \, .
\ees
The Dirichlet boundary condition, on the other hand, can be written in the Mellin space as an integral boundary condition
\be
\label{int_bc}
\widetilde{\Psi}_s (0) = \int_0^\infty \frac{dz}{z}\, e^z g_s (z)   = g_0 (s) (1-2^{1-s}) \Gamma(s) \zeta(s) = 0 \, ,
\ee
and identifies the eigenvalues, $\mathcal{E}_s = i (1/2 -s)$, in terms of the nontrivial zeros of the Riemann zeta function. The result is consistent with the analysis presented in the previous section and the boundary condition~\eqref{bc_zeta}.

The domain of the Hamiltonian can be elaborated via the inner product $ \braket{V| \hat{\widetilde{H}} U} $, where $\ket{U}$ and $\ket{V}$ are arbitrary states (not necessarily the eigenstates) in the Hilbert space. Applying the same steps as above to solve the eigenvalue equation, we expand these arbitrary states in the basis of the eigenstates of the number operator: $\ket{U} = \sum_{n=0}^\infty u_n/n! \ket{n}$, and then we identify the corresponding coefficients in terms of the Mellin transform: $u_n = \int_0^\infty dz \, z^{n-1} u(z)$. We further define $u(z) = s(z) u'(z)$, with $s(z) = z e^{-z}/(1+e^z)$. The state, $\ket{V}$, can be expanded in the same way. As a result of these, the inner product in the Mellin space is given by
\be
\label{self_adjointness}
 \braket{V| \hat{\widetilde{H}} U} = \int_0^\infty dz \int_0^\infty dy \, \frac{I_0(2\sqrt{y z})}{yz} s(z) s(y)  \, v'^*(y) \left( - i z \frac{d}{d z} -\frac{i}{2}  \right) u'(z) \, ,
\ee
where $I_0(2\sqrt{y z})$ is the modified Bessel function of the first kind.

Now, we assume the existence of a similarity transformation such that the transformed Hamiltonian, denoted as $\hat{H}'$, is self-adjoint on the domain specified by an appropriate boundary condition, corresponding to $\widetilde{\Psi}_s (0)=0$, or equivalently  $\Psi_s(0) = 0$ for the introduced Hamiltonian. The similarity transformation manifests itself as a weight function in the Mellin space, which is identified through the mapping, $I_0 (2\sqrt{zy})/(yz) d z \, dy \to \omega(z,y) d z \, dy$. Further elaboration can be found in Appendix~B. Accordingly, we substitute the inner product~\eqref{self_adjointness} with the following
\be
\label{self_adjointness_1}
\braket{V| \hat{\widetilde{H}} U} \to \braket{V| \hat{H}' U} =  \int_0^\infty dz \int_0^\infty dy  \, \omega'(z,y) v'^* (y) \left( - i z \frac{d}{d z} - \frac{i}{2} \right) u'(z) \, ,
\ee
where we define $\omega'(z,y) = \omega(z,y) s(z) s(y)$. Furthermore, all square-integrable functions can be identified in the Mellin space via
\bes
\braket{U|U} = \int_0^\infty dz \int_0^\infty dy  \, \omega '(z,y) u'^* (y) u'(z) \, ,
\ees
which imposes that 
\be
\label{assumption0}
\omega'(z,y) = \omega'^*(y,z) \, .
\ee

After the integration by parts, Eq.~\eqref{self_adjointness_1} can be expressed as
\bes
\braket{V| \hat{H}' U} = BT_z  + \int_0^\infty dz  \int_0^\infty dy \, u'(z)  \left( i z \frac{d}{dz} + \frac{i}{2} \right) \omega'(z,y) v'^* (y) \, ,
\ees
where the boundary term is
\bes
BT_z = \left. - i z u'(z) \int_0^\infty d y \, \omega'(z,y) v'^* (y) \right|_{z=0}^\infty \, .
\ees
Furthermore, we assume the condition
\be
\label{assumption1}
z \frac{d}{d z}  \omega'(z,y) = -  \frac{d}{d y} \left( y \,  \omega'^*(y,z) \right) \, ,
\ee
which suggests $z\, \omega'(z,y)$ as a function of $z/y$. Then, we obtain
\be
\label{self_adjointness_2}
 \braket{V| \hat{H}' U} = BT_z + BT_y  + \int_0^\infty dy  \int_0^\infty dz  \, \omega'^*(y,z) u'(z) \left( i y \frac{d}{dy} + \frac{i}{2} \right) v'^* (y) \, ,
\ee
where the boundary term in the variable $y$ emerges after the integration by parts;
\bes
BT_y = \left. - i y v'^*(y) \int_0^\infty d z \, \omega'^*(y,z) u' (z) \right|_{y=0}^\infty \, .
\ees
Comparing Eqs.~\eqref{self_adjointness_1} and \eqref{self_adjointness_2}, we conclude that
\bes
\braket{V| \hat{H}' U}  = BT + \braket{U| \hat{H}' V}^* \, ,
\ees
where $ BT = BT_z + BT_y $. If the boundary term vanishes, then the transformed Hamiltonian, $\hat{H}'$, becomes formally self-adjoint (Hermitian or symmetric). To establish the self-adjointness of the Hamiltonian, it is necessary to specify a domain such that it matches the adjoint domain, $\text{Dom} \, \hat{H}' = \text{Dom}\, \hat{H}'^*$. This is possible if and only if the bilinear form of the boundary term vanishes identically. Accordingly, we write the boundary term as
\bes 
BT = \left. - i u'(z) z v'^*(z) \, \left( \frac{1}{v'^*(z)} \int_0^\infty dy \, \omega'(z,y) v'^*(y) + \frac{1}{u'(z)} \int_0^\infty dy \, \omega'^*(z,y) u'(y) \right) \right|_{z=0}^\infty \, .
\ees
Then, under the condition of
\be
\label{domain_cond}
 \lim_{z \to 0^+} \frac{1}{u'(z)} \int_0^\infty dy \, \omega'^*(z,y) u'(y) =  \lim_{z \to \infty} \frac{1}{u'(z)} \int_0^\infty dy \, \omega'^*(z,y) u'(y)  =  0 \, ,
\ee
the two domains become the same: $\text{Dom} \, \hat{H}' = \text{Dom}\, \hat{H}'^*$. Thereby, the transformed Hamiltonian becomes self-adjoint on the domain specified by the condition~\eqref{domain_cond}, and hence admits only real eigenvalues. If we further identify the weight function, in accordance with the conditions~\eqref{assumption0} and \eqref{assumption1}, as follows
\be
\label{weight}
\omega'(z,y) = \left( \frac{1/z}{1+ e^{y/z}} + \frac{1/y}{1+ e^{z/y}} \right) \, ,
\ee
then the condition~\eqref{domain_cond} corresponds to the boundary condition, $\Psi_s(0) = 0$, for the eigenfunctions, $g'_s (z) =  s^{-1}(z) \, g_s (z) = g_0 (s)  z^{s-1}$. 

Thus, there exist a similarity transformation of the introduced Hamiltonian~\eqref{initial_ham}, which is self-adjoint on the domain specified by the condition~\eqref{domain_cond}. Moreover, as this condition can be met by the eigenfunctions through the Dirichlet boundary condition, $\Psi_s(0) = 0$, i.e., through the vanishing of the Riemann zeta function, the eigenvalues represent the nontrivial Riemann zeros.

\section*{Discussion} \label{sec_conc}

As a significant observation, replacing the operator $\left( 1 + e^{\hat{x}} \right)^{-1}$ in the Hamiltonian~\eqref{initial_ham} with an arbitrary operator $W(\hat{x})$ leads to the following Dirichlet boundary condition:
\bes
 \Psi_s (0) = \frac{1}{\sqrt{2\pi}} \frac{\sqrt{t}}{1-t} \int_0^\infty dz\, z^{s-1} \,  W(z)  =  \frac{1}{\sqrt{2\pi}} \frac{\sqrt{t}}{1-t} \{\mathcal{M} W \}(s) = 0 \, .
\ees
This implies that our approach can be generalized to a broader class of functions. Specifically, if there exists a similarity transformation of the corresponding Hamiltonian, which is self-adjoint on the domain specified by an appropriate boundary condition consistent with the above Dirichlet boundary condition, then the Mellin transform, $\{\mathcal{M} W \}(s)$, if it exists, can have zeros only on the critical line. Indeed, it is known that the zeros of the Mellin transforms of certain orthogonal polynomials lie only on the critical line~\cite{bump1986riemann,coffey2015mellin}, which is known as the local Riemann hypothesis~\cite{bump2000local}.

To sum up, in this manuscript we have introduced a well-defined Hamiltonian that could possibly satisfy the two stages of the Hilbert-Pólya conjecture; (I) if the RH holds true, its eigenvalues are given by the imaginary parts of the nontrivial Riemann zeros, and (II) it admits only real eigenvalues, as there exists a similarity transformation of the introduced Hamiltonian which is self-adjoint on the domain, where the eigenfunctions satisfy the imposed boundary condition through the vanishing of the Riemann zeta function. If the formal argument presented in this manuscript can be rigorously established, our approach holds the potential to imply the validity of the RH. It is important to note that the second stage requires further rigorous treatment. While we attempted to show the self-adjointness by reformulating the Hamiltonian in the Mellin space, alternative, more rigorous methods could be employed to bypass this analysis.

\begin{acknowledgments}

We acknowledge Michael Berry and Jean Bellissard for valuable discussions.

\end{acknowledgments}

\newpage

\section*{Appendix} \label{appendix}

\subsection{Borel Summation}
\label{app_borel}

The Borel sum of a given series $A(x) = \sum_{n=0}^\infty a_n x^n$ can be defined by
\bes
A(x) (\vec{B}) = \int_0^\infty dt \, e^{-t} \mathcal{B}A(x t) \, ,
\ees
where $\mathcal{B}A(x t) = \sum_{n=0}^\infty a_n  (x t)^n / n! $ is the Borel transform. In general, the Borel sum of a given series can be analytically continued to a larger region than the region where the original sum converges.

We first show that the Borel sum of the series
\bes
s_1(x) = \sum_{n=0}^\infty L_n (z) x^n \, ,
\ees
which converges to $e^{-x z/(1-x)}/(1-x)$ for $|x| < 1$, can be analytically continued to all $\text{Re}(x)< 1$.

Accordingly, the Borel sum of the series $s_1(x)$ can be expressed as
\bes
s_1(x)(\vec{B}) = \int_0^\infty dt \, e^{-t} \sum_{n=0}^\infty \frac{L_n (z)}{n!} (x t)^n \, .
\ees
Using the following integral representation of the Laguerre polynomials
\bes
L_{n} (z) = \frac{e^z}{n!} \int_0^\infty d u \, e^{-u} u^n J_0 (2 \sqrt{z u} ) \, ,
\ees
with $J_m $ being the Bessel function of the first kind, the Borel sum can be written as
\be
\label{borel_sum}
s_1(x)(\vec{B}) = e^z \int_0^\infty d u  \, e^{-u} J_0 (2 \sqrt{z u} ) \int_0^\infty dt \, e^{-t}  \sum_{n=0}^\infty \frac{(t u x)^n}{(n!)^2} \, .
\ee
The sum inside the integral can be identified in terms of the modified Bessel function of the first kind as $I_0 (2\sqrt{t u x})$. After taking the corresponding $t$-integral in Eq.~\eqref{borel_sum}, the Borel sum reads
\bes
 s_1(x)(\vec{B}) = e^z \int_0^\infty d u  \, e^{u(x-1)} J_0 (2 \sqrt{z u} ) \,.
\ees
The $u$-integral converges in the larger region $\text{Re}(x) < 1$ so that it gives an analytic continuation of the original series:
\bes
\\
s_1(x)(\vec{B})  = \frac{e^{-z x /(1-x)}}{1-x} \, , \quad \text{Re}(x) < 1 \,.
\ees

In a similar way, we can show that the Borel sum of the series 
\bes
s_2(x) = \sum_{n=0}^\infty \frac{B_n}{n!} x^n \, ,
\ees
which converges to $x/(1-e^{-x})$ for $0 <|x| < 2 \pi$, can be analytically continued to all $\text{Re}(x)>0$ and $\text{Im}(x)< 2 \pi $.

The Borel sum of the series $s_2(x)$ can be expressed as
\bes
s_2(x)(\vec{B}) = \int_0^\infty dt \, e^{-t} \sum_{n=0}^\infty \frac{B_n}{(n!)^2} (t x)^n  = \int_0^\infty dt \, e^{-t} \left(1 + \frac{1}{2} t x + \sum_{n=1}^\infty \frac{B_{2n}}{(2n!)^2} (t x)^{2n} \right) \, .
\ees
Using the following integral representation of the even Bernoulli numbers
\bes
B_{2n} = 4n (-1)^{n+1} \int_0^\infty du\, \frac{u^{2n-1}}{e^{2\pi u} - 1} \, ,
\ees
the Borel sum can be written as
\be
\label{borel_sum_2}
 s_2(x)(\vec{B}) =  1 + \frac{x}{2} - 4 \int_0^\infty dt \, e^{-t}  \int_0^\infty \frac{du}{u} \frac{1}{e^{2\pi u}-1} \,\sum_{n=1}^\infty  \frac{n (-1)^n}{(2n!)^2} (t x u)^{2n}  \, .
\ee
The sum inside the integral can be identified in terms of the Kelvin functions, $\text{ber}_1$ and $\text{bei}_1$, as 
\bes
 \sum_{n=1}^\infty  \frac{n (-1)^n}{(2n!)^2} (t x u)^{2n} = \frac{\sqrt{t x u}}{2 \sqrt{2}} \left(\text{ber}_1\left(2 \sqrt{t x u}\right)+\text{bei}_1\left(2 \sqrt{t x u}\right)\right) \, .
\ees
Then, the corresponding $t$-integral in Eq.~\eqref{borel_sum_2}, which is given by
\bes
 \int_0^\infty dt \, e^{-t} \frac{\sqrt{t x u}}{2 \sqrt{2}} \left(\text{ber}_1\left(2 \sqrt{t x u}\right)+\text{bei}_1\left(2 \sqrt{t x u}\right)\right) = - \frac{u x}{2} \sin(u x) \, , \quad \text{Re}(x)>0 \, ,
\ees
defines the region where the series is Borel summable. Finally, after performing the remaining $u$-integral, which converges for  $\text{Im}(x)< 2 \pi$, we obtain
\bes
 s_2(x)(\vec{B})  =  1 + \frac{x}{2} + \frac{x \coth \left(x/2 \right)-2}{2}  = \frac{x}{1-e^{-x}} \, , \quad \text{Re}(x)>0 \ \text{and} \ \text{Im}(x)< 2 \pi \, ,
\ees
which gives an analytic continuation of the original series.

\subsection{The inner product and the weight function in the Mellin space}
\label{app_weight_mellin}

The inner product~\eqref{self_adjointness_1} can be expressed as:
\be
\label{inner_prod_app}
\braket{V| \hat{H}' U} = \braket{V| \hat{S}' \hat{\widetilde{H}}  \hat{S}'^{-1} U}  \, .
\ee
After expanding $\hat{S}'^{-1} \ket{U} = \sum_{n=0}^\infty u_n \ket{n}/n!$ and identifying the coefficients in terms of the Mellin transform: $u_n = \int_0^\infty dz \, z^{n-1} u(z)$, the inner product~\eqref{inner_prod_app} can be written as
\bes
\braket{V| \hat{H}' U} = \int_0^\infty dz \int_0^\infty dy \, \omega(z,y) v^*(y) \left( - i z \frac{d}{d z} + \frac{i}{2} - i z - \frac{i z}{1+e^{-z}} \right) u(z) \, ,
\ees
where the weight function in the Mellin space is defined in terms the transformation $\hat{S}'$ as
\be
\omega(z,y) = \sum_{k=0}^\infty \sum_{n=0}^\infty  \frac{y^{k-1}}{k!} \frac{z^{n-1}}{n!} \bra{k}\hat{S}'^\dagger \hat{S'} \ket{n} \, ,
\ee
which reads $\omega(z,y) = I_0 (2\sqrt{z y})/(zy)$ in the trivial case of $\hat{S}' = \hat{I}$. Finally, by defining $u(z) = s(z) u'(z)$ with $s(z) = z e^{-z}/(1+e^z)$, we obtain
\bes
\braket{V| \hat{H}' U} = \int_0^\infty dz \int_0^\infty dy \, \omega'(z,y) v'^*(y) \left( - i z \frac{d}{d z} - \frac{i}{2} \right) u'(z) \, ,
\ees
where $\omega'(z,y) = \omega(z,y) s(z) s(y)$.

We would also like to emphasize here that the transformed eigenfunctions are square-integrable for the nontrivial Riemann zeros, $\rho$, in the following sense:
\bals
\braket{\Psi'_{\rho}|\Psi'_{\rho'}} & = \int_0^\infty d z \int_0^\infty d y \,  \omega'(z,y) g'^*_\rho (y) g'_{\rho'} (z) \\
& = g_0^* (\rho) g_0 (\rho')  \int_0^\infty d z\, z^{\rho^*+\rho' -2} \int_0^\infty d y\, \frac{y^{\rho^*-1} + y^{\rho'-1}}{1+e^y} \\
& = \delta_{\rho \rho'} \, |g_0 (\rho)|^2 \lim_{\epsilon \to 0} \int_0^\infty d z\, z^{2 \epsilon - 1} \left( F(\rho^* + \epsilon) + F(\rho + \epsilon) \right) \\
& =  \delta_{\rho \rho'} \, |g_0 (\rho)|^2 \text{Re}(F'(\rho))  \, ,
\eals
where $F(s) = (1-2^{1-s}) \Gamma(s)\zeta(s)$ and we employ $\int_0^\infty dz \, \delta(z) = 1$.

%We would also like to emphasize how the inner product can be expressed for the transformed eigenfunctions:
%\bals
%\braket{\Psi'_{s}|\Psi'_{s'}} & = \int_0^\infty d z \int_0^\infty d y \,  \omega'(z,y) g'^*_s (y) g'_{s'} (z) = \lim_{\epsilon \to 0^+} |g_0|^2 \int_\epsilon^\infty d z\, z^{s^*+s' -2} \int_0^\infty d y\, \frac{y^{s^*-1} + y^{s'-1}}{1+e^y} \\
%& = |g_0|^2 2 \pi \delta (\mathcal{E}_s - \mathcal{E}_{s'}) \text{Re}\left((1-2^{1-s})\Gamma(s) \zeta(s) \right) \, .
%\eals
%It would be intriguing to investigate whether this result can be interpreted as a Kronecker delta, ensuring the square-integrability of the eigenfunctions with respect to the nontrivial Riemann zeros.

% \newpage

\bibliography{bib_zeta.bib}

\begin{thebibliography}{25}%
\makeatletter
\providecommand \@ifxundefined [1]{%
 \@ifx{#1\undefined}
}%
\providecommand \@ifnum [1]{%
 \ifnum #1\expandafter \@firstoftwo
 \else \expandafter \@secondoftwo
 \fi
}%
\providecommand \@ifx [1]{%
 \ifx #1\expandafter \@firstoftwo
 \else \expandafter \@secondoftwo
 \fi
}%
\providecommand \natexlab [1]{#1}%
\providecommand \enquote  [1]{``#1''}%
\providecommand \bibnamefont  [1]{#1}%
\providecommand \bibfnamefont [1]{#1}%
\providecommand \citenamefont [1]{#1}%
\providecommand \href@noop [0]{\@secondoftwo}%
\providecommand \href [0]{\begingroup \@sanitize@url \@href}%
\providecommand \@href[1]{\@@startlink{#1}\@@href}%
\providecommand \@@href[1]{\endgroup#1\@@endlink}%
\providecommand \@sanitize@url [0]{\catcode `\\12\catcode `\$12\catcode
  `\&12\catcode `\#12\catcode `\^12\catcode `\_12\catcode `\%12\relax}%
\providecommand \@@startlink[1]{}%
\providecommand \@@endlink[0]{}%
\providecommand \url  [0]{\begingroup\@sanitize@url \@url }%
\providecommand \@url [1]{\endgroup\@href {#1}{\urlprefix }}%
\providecommand \urlprefix  [0]{URL }%
\providecommand \Eprint [0]{\href }%
\providecommand \doibase [0]{http://dx.doi.org/}%
\providecommand \selectlanguage [0]{\@gobble}%
\providecommand \bibinfo  [0]{\@secondoftwo}%
\providecommand \bibfield  [0]{\@secondoftwo}%
\providecommand \translation [1]{[#1]}%
\providecommand \BibitemOpen [0]{}%
\providecommand \bibitemStop [0]{}%
\providecommand \bibitemNoStop [0]{.\EOS\space}%
\providecommand \EOS [0]{\spacefactor3000\relax}%
\providecommand \BibitemShut  [1]{\csname bibitem#1\endcsname}%
\let\auto@bib@innerbib\@empty
%</preamble>
\bibitem [{\citenamefont {Titchmarsh}\ and\ \citenamefont
  {Heath-Brown}(1986)}]{titchmarsh1986theory}%
  \BibitemOpen
  \bibfield  {author} {\bibinfo {author} {\bibfnamefont {E.~C.}\ \bibnamefont
  {Titchmarsh}}\ and\ \bibinfo {author} {\bibfnamefont {D.~R.}\ \bibnamefont
  {Heath-Brown}},\ }\href@noop {} {\emph {\bibinfo {title} {The theory of the
  Riemann zeta-function}}}\ (\bibinfo  {publisher} {Oxford university press},\
  \bibinfo {year} {1986})\BibitemShut {NoStop}%
\bibitem [{\citenamefont {Gourdon}\ and\ \citenamefont
  {Demichel}(2004)}]{gourdon20041013}%
  \BibitemOpen
  \bibfield  {author} {\bibinfo {author} {\bibfnamefont {X.}~\bibnamefont
  {Gourdon}}\ and\ \bibinfo {author} {\bibfnamefont {P.}~\bibnamefont
  {Demichel}},\ }\href@noop {} {\enquote {\bibinfo {title} {The $10^{13}$ first
  zeros of the riemann zeta function, and zeros computation at very large
  height},}\ } (\bibinfo {year} {2004})\BibitemShut {NoStop}%
\bibitem [{\citenamefont {Selberg}(1956)}]{selberg1956harmonic}%
  \BibitemOpen
  \bibfield  {author} {\bibinfo {author} {\bibfnamefont {A.}~\bibnamefont
  {Selberg}},\ }\href@noop {} {\bibfield  {journal} {\bibinfo  {journal} {J.
  Indian Math. Soc.}\ }\textbf {\bibinfo {volume} {20}},\ \bibinfo {pages} {47}
  (\bibinfo {year} {1956})}\BibitemShut {NoStop}%
\bibitem [{\citenamefont {Montgomery}(1973)}]{montgomery1973pair}%
  \BibitemOpen
  \bibfield  {author} {\bibinfo {author} {\bibfnamefont {H.~L.}\ \bibnamefont
  {Montgomery}},\ }in\ \href@noop {} {\emph {\bibinfo {booktitle} {Proc. Symp.
  Pure Math}}},\ Vol.~\bibinfo {volume} {24}\ (\bibinfo {year} {1973})\ pp.\
  \bibinfo {pages} {181--193}\BibitemShut {NoStop}%
\bibitem [{\citenamefont {Odlyzko}(1987)}]{odlyzko1987distribution}%
  \BibitemOpen
  \bibfield  {author} {\bibinfo {author} {\bibfnamefont {A.~M.}\ \bibnamefont
  {Odlyzko}},\ }\href@noop {} {\bibfield  {journal} {\bibinfo  {journal}
  {Mathematics of Computation}\ }\textbf {\bibinfo {volume} {48}},\ \bibinfo
  {pages} {273} (\bibinfo {year} {1987})}\BibitemShut {NoStop}%
\bibitem [{\citenamefont {Okubo}(1998)}]{okubo1998lorentz}%
  \BibitemOpen
  \bibfield  {author} {\bibinfo {author} {\bibfnamefont {S.}~\bibnamefont
  {Okubo}},\ }\href@noop {} {\bibfield  {journal} {\bibinfo  {journal} {Journal
  of Physics A: Mathematical and General}\ }\textbf {\bibinfo {volume} {31}},\
  \bibinfo {pages} {1049} (\bibinfo {year} {1998})}\BibitemShut {NoStop}%
\bibitem [{\citenamefont {Connes}(1999)}]{connes1999trace}%
  \BibitemOpen
  \bibfield  {author} {\bibinfo {author} {\bibfnamefont {A.}~\bibnamefont
  {Connes}},\ }\href@noop {} {\bibfield  {journal} {\bibinfo  {journal}
  {Selecta Mathematica}\ }\textbf {\bibinfo {volume} {5}},\ \bibinfo {pages}
  {29} (\bibinfo {year} {1999})}\BibitemShut {NoStop}%
\bibitem [{\citenamefont {Berry}\ and\ \citenamefont
  {Keating}(1999{\natexlab{a}})}]{berry1999h}%
  \BibitemOpen
  \bibfield  {author} {\bibinfo {author} {\bibfnamefont {M.~V.}\ \bibnamefont
  {Berry}}\ and\ \bibinfo {author} {\bibfnamefont {J.~P.}\ \bibnamefont
  {Keating}},\ }in\ \href@noop {} {\emph {\bibinfo {booktitle} {Supersymmetry
  and Trace Formulae}}}\ (\bibinfo  {publisher} {Springer},\ \bibinfo {year}
  {1999})\ pp.\ \bibinfo {pages} {355--367}\BibitemShut {NoStop}%
\bibitem [{\citenamefont {Berry}\ and\ \citenamefont
  {Keating}(1999{\natexlab{b}})}]{berry1999riemann}%
  \BibitemOpen
  \bibfield  {author} {\bibinfo {author} {\bibfnamefont {M.~V.}\ \bibnamefont
  {Berry}}\ and\ \bibinfo {author} {\bibfnamefont {J.~P.}\ \bibnamefont
  {Keating}},\ }\href@noop {} {\bibfield  {journal} {\bibinfo  {journal} {SIAM
  review}\ }\textbf {\bibinfo {volume} {41}},\ \bibinfo {pages} {236} (\bibinfo
  {year} {1999}{\natexlab{b}})}\BibitemShut {NoStop}%
\bibitem [{\citenamefont {Sierra}(2007)}]{sierra2007h}%
  \BibitemOpen
  \bibfield  {author} {\bibinfo {author} {\bibfnamefont {G.}~\bibnamefont
  {Sierra}},\ }\href@noop {} {\bibfield  {journal} {\bibinfo  {journal}
  {Nuclear Physics B}\ }\textbf {\bibinfo {volume} {776}},\ \bibinfo {pages}
  {327} (\bibinfo {year} {2007})}\BibitemShut {NoStop}%
\bibitem [{\citenamefont {Sierra}(2019)}]{sierra2019riemann}%
  \BibitemOpen
  \bibfield  {author} {\bibinfo {author} {\bibfnamefont {G.}~\bibnamefont
  {Sierra}},\ }\href@noop {} {\bibfield  {journal} {\bibinfo  {journal}
  {Symmetry}\ }\textbf {\bibinfo {volume} {11}},\ \bibinfo {pages} {494}
  (\bibinfo {year} {2019})}\BibitemShut {NoStop}%
\bibitem [{\citenamefont {Bender}\ \emph {et~al.}(2017)\citenamefont {Bender},
  \citenamefont {Brody},\ and\ \citenamefont
  {M{\"u}ller}}]{bender2017hamiltonian}%
  \BibitemOpen
  \bibfield  {author} {\bibinfo {author} {\bibfnamefont {C.~M.}\ \bibnamefont
  {Bender}}, \bibinfo {author} {\bibfnamefont {D.~C.}\ \bibnamefont {Brody}}, \
  and\ \bibinfo {author} {\bibfnamefont {M.~P.}\ \bibnamefont {M{\"u}ller}},\
  }\href@noop {} {\bibfield  {journal} {\bibinfo  {journal} {Physical Review
  Letters}\ }\textbf {\bibinfo {volume} {118}},\ \bibinfo {pages} {130201}
  (\bibinfo {year} {2017})}\BibitemShut {NoStop}%
\bibitem [{\citenamefont {Bellissard}(2017)}]{bellissard2017comment}%
  \BibitemOpen
  \bibfield  {author} {\bibinfo {author} {\bibfnamefont {J.~V.}\ \bibnamefont
  {Bellissard}},\ }\href@noop {} {\bibfield  {journal} {\bibinfo  {journal}
  {arXiv preprint arXiv:1704.02644}\ } (\bibinfo {year} {2017})}\BibitemShut
  {NoStop}%
\bibitem [{\citenamefont {Das}\ and\ \citenamefont
  {Kalauni}(2019)}]{das2019supersymmetry}%
  \BibitemOpen
  \bibfield  {author} {\bibinfo {author} {\bibfnamefont {A.}~\bibnamefont
  {Das}}\ and\ \bibinfo {author} {\bibfnamefont {P.}~\bibnamefont {Kalauni}},\
  }\href@noop {} {\bibfield  {journal} {\bibinfo  {journal} {Physics Letters
  B}\ }\textbf {\bibinfo {volume} {791}},\ \bibinfo {pages} {265} (\bibinfo
  {year} {2019})}\BibitemShut {NoStop}%
\bibitem [{\citenamefont {Bonneau}\ \emph {et~al.}(2001)\citenamefont
  {Bonneau}, \citenamefont {Faraut},\ and\ \citenamefont
  {Valent}}]{bonneau2001self}%
  \BibitemOpen
  \bibfield  {author} {\bibinfo {author} {\bibfnamefont {G.}~\bibnamefont
  {Bonneau}}, \bibinfo {author} {\bibfnamefont {J.}~\bibnamefont {Faraut}}, \
  and\ \bibinfo {author} {\bibfnamefont {G.}~\bibnamefont {Valent}},\
  }\href@noop {} {\bibfield  {journal} {\bibinfo  {journal} {American Journal
  of physics}\ }\textbf {\bibinfo {volume} {69}},\ \bibinfo {pages} {322}
  (\bibinfo {year} {2001})}\BibitemShut {NoStop}%
\bibitem [{\citenamefont {Faris}(2006)}]{faris2006self}%
  \BibitemOpen
  \bibfield  {author} {\bibinfo {author} {\bibfnamefont {W.~G.}\ \bibnamefont
  {Faris}},\ }\href@noop {} {\emph {\bibinfo {title} {Self-adjoint
  operators}}},\ Vol.\ \bibinfo {volume} {433}\ (\bibinfo  {publisher}
  {Springer},\ \bibinfo {year} {2006})\BibitemShut {NoStop}%
\bibitem [{\citenamefont {Twamley}\ and\ \citenamefont
  {Milburn}(2006)}]{twamley2006quantum}%
  \BibitemOpen
  \bibfield  {author} {\bibinfo {author} {\bibfnamefont {J.}~\bibnamefont
  {Twamley}}\ and\ \bibinfo {author} {\bibfnamefont {G.}~\bibnamefont
  {Milburn}},\ }\href@noop {} {\bibfield  {journal} {\bibinfo  {journal} {New
  Journal of Physics}\ }\textbf {\bibinfo {volume} {8}},\ \bibinfo {pages}
  {328} (\bibinfo {year} {2006})}\BibitemShut {NoStop}%
\bibitem [{\citenamefont {Reed}\ and\ \citenamefont
  {Simon}(1981)}]{reed1981functional}%
  \BibitemOpen
  \bibfield  {author} {\bibinfo {author} {\bibfnamefont {M.}~\bibnamefont
  {Reed}}\ and\ \bibinfo {author} {\bibfnamefont {B.}~\bibnamefont {Simon}},\
  }\href@noop {} {\emph {\bibinfo {title} {I: Functional analysis}}},\
  Vol.~\bibinfo {volume} {1}\ (\bibinfo  {publisher} {Academic press},\
  \bibinfo {year} {1981})\BibitemShut {NoStop}%
\bibitem [{\citenamefont {Arfken}\ \emph {et~al.}(2011)\citenamefont {Arfken},
  \citenamefont {Weber},\ and\ \citenamefont
  {Harris}}]{arfken2011mathematical}%
  \BibitemOpen
  \bibfield  {author} {\bibinfo {author} {\bibfnamefont {G.~B.}\ \bibnamefont
  {Arfken}}, \bibinfo {author} {\bibfnamefont {H.~J.}\ \bibnamefont {Weber}}, \
  and\ \bibinfo {author} {\bibfnamefont {F.~E.}\ \bibnamefont {Harris}},\
  }\href@noop {} {\emph {\bibinfo {title} {Mathematical methods for physicists:
  a comprehensive guide}}}\ (\bibinfo  {publisher} {Academic press},\ \bibinfo
  {year} {2011})\BibitemShut {NoStop}%
\bibitem [{\citenamefont {Teschl}(2014)}]{teschl2014mathematical}%
  \BibitemOpen
  \bibfield  {author} {\bibinfo {author} {\bibfnamefont {G.}~\bibnamefont
  {Teschl}},\ }\href@noop {} {\emph {\bibinfo {title} {Mathematical methods in
  quantum mechanics}}},\ Vol.\ \bibinfo {volume} {157}\ (\bibinfo  {publisher}
  {American Mathematical Soc.},\ \bibinfo {year} {2014})\BibitemShut {NoStop}%
\bibitem [{\citenamefont {Perelomov}(1977)}]{perelomov1977generalized}%
  \BibitemOpen
  \bibfield  {author} {\bibinfo {author} {\bibfnamefont {A.~M.}\ \bibnamefont
  {Perelomov}},\ }\href@noop {} {\bibfield  {journal} {\bibinfo  {journal}
  {Soviet Physics Uspekhi}\ }\textbf {\bibinfo {volume} {20}},\ \bibinfo
  {pages} {703} (\bibinfo {year} {1977})}\BibitemShut {NoStop}%
\bibitem [{\citenamefont {Barut}\ and\ \citenamefont
  {Girardello}(1971)}]{barut1971new}%
  \BibitemOpen
  \bibfield  {author} {\bibinfo {author} {\bibfnamefont {A.}~\bibnamefont
  {Barut}}\ and\ \bibinfo {author} {\bibfnamefont {L.}~\bibnamefont
  {Girardello}},\ }\href@noop {} {\bibfield  {journal} {\bibinfo  {journal}
  {Communications in Mathematical Physics}\ }\textbf {\bibinfo {volume} {21}},\
  \bibinfo {pages} {41} (\bibinfo {year} {1971})}\BibitemShut {NoStop}%
\bibitem [{\citenamefont {Bump}\ and\ \citenamefont
  {Ng}(1986)}]{bump1986riemann}%
  \BibitemOpen
  \bibfield  {author} {\bibinfo {author} {\bibfnamefont {D.}~\bibnamefont
  {Bump}}\ and\ \bibinfo {author} {\bibfnamefont {E.~K.~S.}\ \bibnamefont
  {Ng}},\ }\href@noop {} {\bibfield  {journal} {\bibinfo  {journal}
  {Mathematische Zeitschrift}\ }\textbf {\bibinfo {volume} {192}},\ \bibinfo
  {pages} {195} (\bibinfo {year} {1986})}\BibitemShut {NoStop}%
\bibitem [{\citenamefont {Coffey}\ and\ \citenamefont
  {Lettington}(2015)}]{coffey2015mellin}%
  \BibitemOpen
  \bibfield  {author} {\bibinfo {author} {\bibfnamefont {M.~W.}\ \bibnamefont
  {Coffey}}\ and\ \bibinfo {author} {\bibfnamefont {M.~C.}\ \bibnamefont
  {Lettington}},\ }\href@noop {} {\bibfield  {journal} {\bibinfo  {journal}
  {Journal of Number Theory}\ }\textbf {\bibinfo {volume} {148}},\ \bibinfo
  {pages} {507} (\bibinfo {year} {2015})}\BibitemShut {NoStop}%
\bibitem [{\citenamefont {Bump}\ \emph {et~al.}(2000)\citenamefont {Bump},
  \citenamefont {Choi}, \citenamefont {Kurlberg},\ and\ \citenamefont
  {Vaaler}}]{bump2000local}%
  \BibitemOpen
  \bibfield  {author} {\bibinfo {author} {\bibfnamefont {D.}~\bibnamefont
  {Bump}}, \bibinfo {author} {\bibfnamefont {K.-K.}\ \bibnamefont {Choi}},
  \bibinfo {author} {\bibfnamefont {P.}~\bibnamefont {Kurlberg}}, \ and\
  \bibinfo {author} {\bibfnamefont {J.}~\bibnamefont {Vaaler}},\ }\href@noop {}
  {\bibfield  {journal} {\bibinfo  {journal} {Mathematische Zeitschrift}\
  }\textbf {\bibinfo {volume} {233}},\ \bibinfo {pages} {1} (\bibinfo {year}
  {2000})}\BibitemShut {NoStop}%
\end{thebibliography}%

\end{document}